# Hyperbolic Metamaterial Resonator-Antenna Scheme for Large, Broadband Emission Enhancement and Single Photon Collection

Faraz A Inam[1,2*], Nadeem Ahmed[1], Michael J. Steel[3] and Stefania Castelletto[4]

*[1]Dept. of Physics, Aligarh Muslim University, Aligarh, U.P. 202002, India*
*[2]Dept. of Condensed Matter Physics and Material Science, Tata Institute of Fundamental Research, Mumbai 400005, India*
*[3]MQ Photonics Research Centre, Dept. of Physics and Astronomy, Macquarie University, Sydney, NSW 2109, Australia*
*[4]School of Engineering, RMIT University, Melbourne, Victoria 3001, Australia*
*\*Author e-mail address: faraz.inam.phy@amu.ac.in*

**Abstract:**

We model the broadband enhancement of single-photon emission from color centres in silicon carbide nanocrystals coupled to a planar hyperbolic metamaterial (HMM) resonator. The design is based on positioning the single photon emitters within the HMM resonator, made of a dielectric index-matched with silicon-carbide material. The broadband response results from the successive resonance peaks of the lossy Fabry-Perot structure modes arising within the high-index HMM cavity. To capture this broadband enhancement in the single photon emitter's spontaneous emission, we placed a simple gold based cylindrical antenna on top of the HMM resonator. We analyzed the performance of this HMM coupled antenna structure in terms of the Purcell enhancement, quantum efficiency, collection efficiency and overall collected photon rate. For perpendicular dipole orientation relative to the interface, the HMM coupled antenna resonator leads to a significantly large spontaneous emission enhancement with Purcell factor of the order of 250 along with a very high average total collected photon rate (CPR) of about 30 over a broad emission spectrum (700 nm – 1000 nm). The peak CPR increases to about 80 at 900 nm, corresponding to the emission of silicon-carbide quantum emitters. This is a state-of-the art improvement considering the previous computational designs have reported a maximum average CPR of 25 across the nitrogen-vacancy centre emission spectrum, 600 nm to 800 nm with the highest value being about 40 at 650 nm.

***Keyworks:*** *Hyperbolic metamaterial resonator, Single-photon sources, Silicon Carbide emitters, nitrogen vacancy centres, broadband enhancement, single photon antenna.*

# I. INTRODUCTION.

Solid-state quantum emitters operating at room temperature [1,2], and their integration in photonic networks are central in the development of quantum information processing. A variety of materials have emerged as an alternative platform to host color centres that could operate as room-temperature single photon sources (SPSs) [3,4] and at the same time, as coherent spin quantum bits with optical read-out [5,6]. This category of quantum emitters includes diamond color centres [7], zinc-oxide based (ZnO) color centres [8], emitters in silicon-carbide (SiC) [3], GaN [9], and 2D materials such as boron nitride [4]. While all these platforms are of great interest for future integrated on-chip quantum photonics at room temperature, as they all have their own set of interesting properties, only diamond and SiC have been proved to host SPSs and spin qubits with optical read-out at the same time [5,6]. In particular SiC can host bright SPSs in the visible [10] that may be efficiently electrically driven using off-the-shelf fabrication methods [11]. Further, SiC possesses many color centres with high spin state [3], that can be used as quantum bits with the longest coherence time in solids, due to the absence of spin-orbit coupling and nuclear spin bath decoupling [12]. Additionally SiC is a CMOS compatible material [13] with a large variety of nanofabrication methods that includes laser micromachining/deposition and ablation [14], making it ideally placed for device fabrication and on-chip integration of quantum systems [15]. These quantum emitters' spectra range from the visible region to the near infrared (up to ~1100nm), the latter emitters being much dimmer [6] and thus most in need of radiative spontaneous emission (SE) enhancement for their application in magnetic sensing [16] and spin nanoscopy [17].

As color centres are point objects emitting radiation as a point dipole with a slow decay rate (nanosecond scale), one of the main aims is to increase their decay rate and direct their emission modes. The typical approach followed is to couple them with microcavities or nanocavities with high quality factors or small mode volume to increase the enhancement on a specific emission mode and couple them to waveguides to extract the photons [2,18]. For optical micro and nanocavities, large Purcell enhancement ($F_p$) is limited to narrow linewidths. The other approach is to build a single photon antenna [19] based on plasmonic resonators [20]. Here the quality factor is low, thus enhancement can occur over the entire color center spectrum (100 nm), while high confinement can be achieved. A major disadvantage of plasmonic antennas is in relying on materials with plasmonic resonance which entails plasmonic losses [21]. More recently a class of Hyperbolic Metamaterials (HMM) [22], metal-dieletric multilayers, have been proposed to enhance the SE of SPSs due to their in-principle indefinitely large photonic density of states and have opened new avenues in their use for quantum optics applications, including SPSs. HMMs have a highly anisotropic dielectric tensor, with dielectric properties in some directions and metallic properties in another direction. This provides the emitting dipole with asymptotically directed large momentum LDOS modes (high-$k$ modes with unbounded magnitude of $K_x$ and $K_z$ in the lossless limit). This enables high emission directionality, a large increase of LDOS and of SE rate, with overall high quantum yield due to a larger ratio of radiated to dissipated power as compared to plasmonic antennas [23]. All these properties are maintained over a 100 to 200 nm spectral region.

The collection enhancement of quantum dots SE has been achieved when positioned within HMM resonators [24], while for the nitrogen vacancy centre (NV), the nanodiamonds were positioned on top of the HMM resonators [25]. This last case has achieved an enhancement of SE of a factor of 3, relative to the emission on a cover glass. Further by introducing a metalloid as metamaterial rather than gold, such as TiN, with lower dissipative losses, a SE enhancement close to 10 for a perpendicular dipole was predicted (5.5 for the average dipole orientation) with an experimental average enhancement of a factor of 4 [26].

The effectiveness of the various out-coupling schemes can be quantified in terms of the overall enhancement in the collected photon rate (CPR) [27]. The CPR is governed by both the Purcell enhancement ($F_p$) and collection efficiency (CE), defined as the spatial overlap of the emission with the numerical aperture (NA) of the collection objective, the CPR being the product of the above two factors ($CPR \propto F_p \cdot CE$) [27]. This means that even if a perfect dielectric antenna is able to achieve perfect collection with a CE of unity but with negligible Purcell enhancement, ($F_p \sim 1$), it results in a CPR of 1. Present state of the art collection schemes for NV centre emission using anti-reflection coated solid-immersion lens reports a CE ~ 0.4 with $F_p$ ~ 1 [28,29] and dielectric grating achieving a CE ~ 0.7 and $F_p$ ~1 [30,31]. Simulated designs of metal antennas [32] have been shown to achieve a $F_p$ ~ 50 and CE ~ 0.3 in a bandwidth of about 50 nm with an average CPR of just about 2-3 over the full NV spectrum. A recent simulated study of metal-dielectric nano-antenna have reported the highest CPR of about 25 across the NV spectrum [27].

In this work, we study and design a HMM resonator based on only five alternating layers of gold and ZnS (previously 16 layers were used [26]) for non-resonant broad-band enhancement of SE of SiC emitters [33]. This structure would be much simpler to fabricate experimentally and very much preserves the hyperbolicity condition ($\epsilon_\perp \cdot \epsilon_\parallel < 0$) in the studied frequency range of 650 nm to 1000 nm. The emitter is positioned in the centre of the resonator rather than on the top. It is thus the first broadband design specifically targeting emitters, such as the carbon antisite vacancy pair [10] and other visible SiC surface quantum emitters [11], the silicon vacancy in SiC [6], and it could be as well applied to NV centres in diamond. The choice of ZnS as dielectric is based on its facile deposition methods with its wide use in thin-film growth [34] and most importantly its refractive index (2.4) being similar to SiC (2.5) and diamond (2.4). Emitters in high index dielectric materials like diamond/SiC generally suffer from low emission rates due to poor index matching at the diamond-air interface [35]. Various methods have been explored to efficiently out-couple the emission from these centers by placing them in an environment of higher refractive index materials [36]. A ZnS resonator with its index matched with both diamond and SiC will be an ideal candidate for the out-coupling of the emission from these host centres. This resonator will be useful in coupling of nanocrystals of SiC containing the red or near infrared emitters [6] or nanodiamonds containing NV centres. The use of SiC thin layers achieved by pulsed laser deposition [37] as a dielectric in the HMM resonators, being the ultimate goal to apply this design to emitters directly embedded in the dielectric constituent [38] of the HMM resonator. This design provides large Purcell factor enhancements ($\Gamma_{HMM}/\Gamma_{vacuum}$) of about 300 at both visible (680 nm, corresponding to the peak of diamond NV centre emission) and near infrared emission (900 nm, corresponding to the emission of SiC nanoparticles) for perpendicular dipole orientation. Further by using a simple Au based cylindrical antenna on top of the HMM we are able to achieve an average collection efficiency of about 0.15. This gives us a very high collected photon rate ($CPR \propto F_p \cdot CE$) of about 30 in the broad spectrum range of 700 nm to 1000 nm.

## II. THEORETICAL BACKGROUND:

The spontaneous emission rate $\Gamma$ of a quantum emitter can be accelerated by tuning the electromagnetic mode environment in its vicinity through a factor known as the local density of optical states (LDOS), $\rho(\mathbf{r}, \omega)$. The LDOS in terms of the Green function is defined as [19,39]

$$\rho(\mathbf{r}, \omega) = (6\omega/\pi c^2)[\hat{\mathbf{d}}^T \cdot \operatorname{Im} G(\omega; \mathbf{r}, \mathbf{r}) \cdot \hat{\mathbf{d}}] \qquad (1)$$

for the emitter source located at **r** and oriented along the unit vector **d̂**. Here $\omega$ is the transition frequency and $G$ is the dyadic Green function signifying the interaction of the radiated electric field with the emitter at the source point.

The enhancement of the spontaneous emission by an environment with high LDOS is known as the Purcell effect, which is quantified in terms of the Purcell factor ($F_p$), i.e. the spontaneous emission decay rate in such an environment compared to that in vacuum. A resonant structure provides an emitter with enhanced LDOS for its emitted photon to couple to when its emission frequency matches with the frequency of the optical modes supported by the structure. In plasmonics, large electric field in the vicinity of the metallic nano-structures significantly enhances the LDOS over the broad scattering spectrum of the metallic nano-structures. Plasmonic structures however provide enhanced LDOS mostly in the visible region of the spectrum as the resonant plasma frequency of the metals lies in this range. A hyperbolic meta-material (HMM) comprises of sub-wavelength scale periodic metal-dielectric layers in a specific direction [22]. This provides anisotropy in the dispersion profile resulting in anisotropy in the LDOS for different transition dipole orientations as well as divergence of the LDOS [22]. In HMMs the modes have higher momentum than conventional surface plasmon polaritons, even out of resonance.

From the layers' permittivity and the exact fill fraction of metal in the structure the components of the effective anisotropic permittivity can be calculated using effective medium theory [40]. The average index for the multilayered HMM can be calculated using the relation [22]:

$$\epsilon_\perp = \frac{\epsilon_m d_m + \epsilon_d d_d}{d_m + d_d}, \qquad \frac{1}{\epsilon_\parallel} = \frac{d_m/\epsilon_m + d_d/\epsilon_d}{d_m + d_d} \qquad (2)$$

where $\epsilon_\perp, \epsilon_\parallel$ are the dielectric components in perpendicular and parallel directions respectively and $d_m, d_d$ are the thickness of metal and dielectric layers respectively. In HMM the effective permittivity components $\epsilon_\parallel$ parallel and perpendicular $\epsilon_\perp$ to the interface have opposite signs ($\epsilon_\perp \cdot \epsilon_\parallel < 0$), which lends a hyperboloid shape to the iso-frequency surface (IFC), according to the dispersion relation equation $\frac{k^2_\parallel}{\epsilon_\parallel} + \frac{k^2_\perp}{\epsilon_\perp} = \frac{\omega^2}{c^2}$. In the case of our ZnS/Au HMM structure (described in the next section), the above hyperbolicity condition is well displayed within our studied frequency range of 650 - 1000 nm (Fig.1).

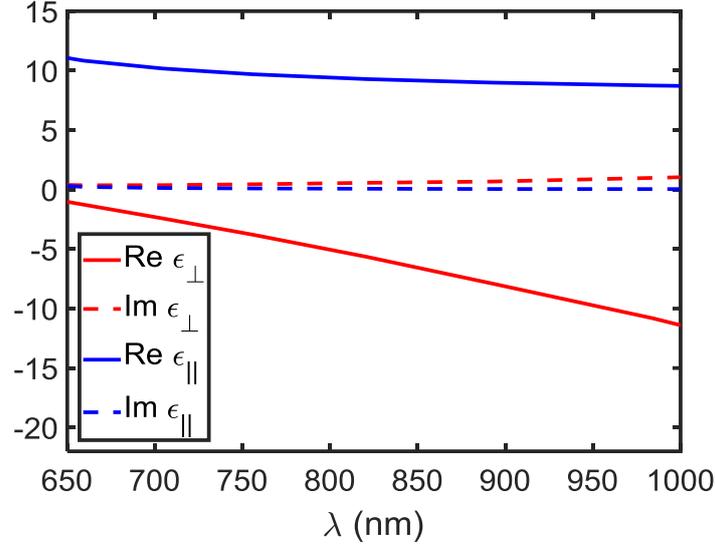

Figure 1. The effective permittivity $\epsilon_\perp$ (red) and $\epsilon_\parallel$ (blue) of our ZnS/Au HMM structure as a function of wavelength.

There are two types of resonant modes arising in a HMM structure. The first is that of the plasmonic modes resulting from the alternate sub-wavelength metal layers. The second type are the structure modes as the HMM structure acts as a nanoscale lossy cavity. An interplay of these two type of modes results in the broadband enhancement of the LDOS seen by the transition dipole when placed inside the HMM [41].

### III. RESULTS AND DISCUSSION:

We carried out calculations using the Radio Frequency (RF) module of the COMSOL Multiphysics suite. To test the validity of our calculations we compared our results with the analytical results of a dipole emission near a metal film (Drexhage experiment) [42]. Figure 2 shows the good agreement of our calculations with the analytical results. In this model we have a dipole emitter positioned in air and 10 nm away from the gold (Au) surface (Fig. 2a).

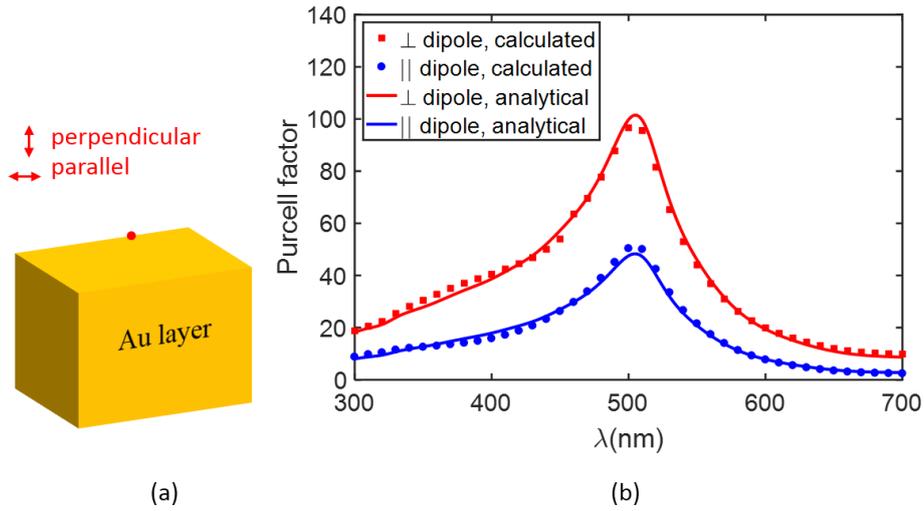

(a)                     (b)

Figure 2. (a) Schematic of a dipole emitter placed on top of the gold (Au) surface. The separation between the dipole and Au surface is considered as 10 nm. (b) Calculated and analytical Purcell factor ($\Gamma/\Gamma_{vacuum}$) for the above considered case of dipole emission above the Au surface.

For this HMM calculation, we have considered quantum emitters based on silicon-carbide (SiC) nano-particles.

**Purcell Factor/Emission Rate Calculations:**

We start the design of HMM structure by encapsulating the nano-spheres of silicon-carbide of diameter 40 nm inside a Poly Vinyl Alcohol (PVA) matrix of thickness 50 nm. The PVA layer is encapsulated on both sides by gold (Au) metal layers of 30 nm thickness. To provide the hyperbolicity to this structure we place a low index dielectric layer of PVA of the same thickness 30 nm above and below the Au layers. This arrangement gives a simple five-layer structure of an HMM. The quantum emitter is placed at the centre of the SiC sphere. The HMM structure is placed on a glass substrate. From this simple low-index PVA/Au HMM, we build our HMM structure by sequentially replacing the PVA layer with the high indexed layer of ZnS (Fig. 3). The ZnS layer has its index matched with the SiC nanoparticle and acts as an effective resonator for enhanced emission from SiC centres.

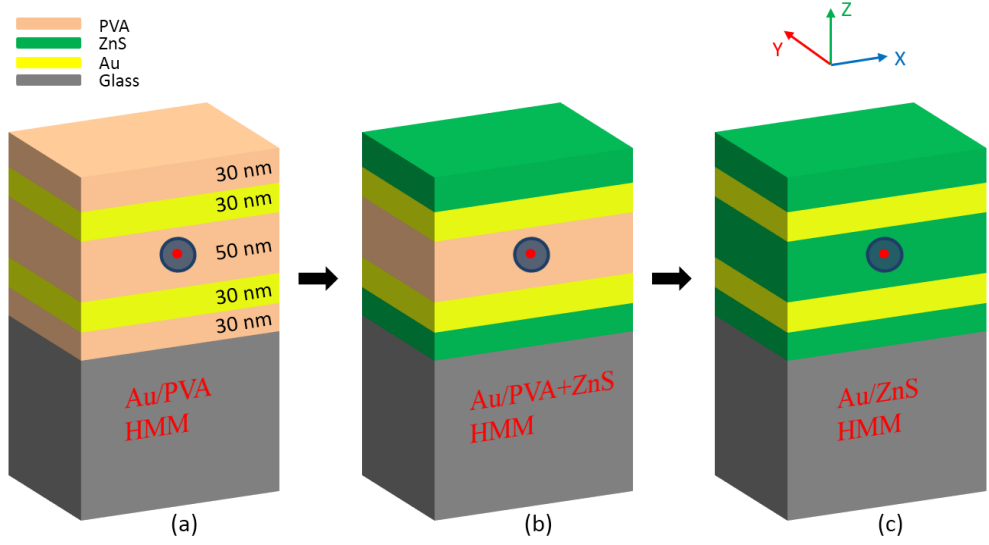

Figure 3. Schematic images showing the design for Au/ZnS HMM starting from a simple Au/PVA counterpart. (a) Au/PVA HMM, (b) Au/PVA+ZnS HMM and (c) Au/ZnS HMM.

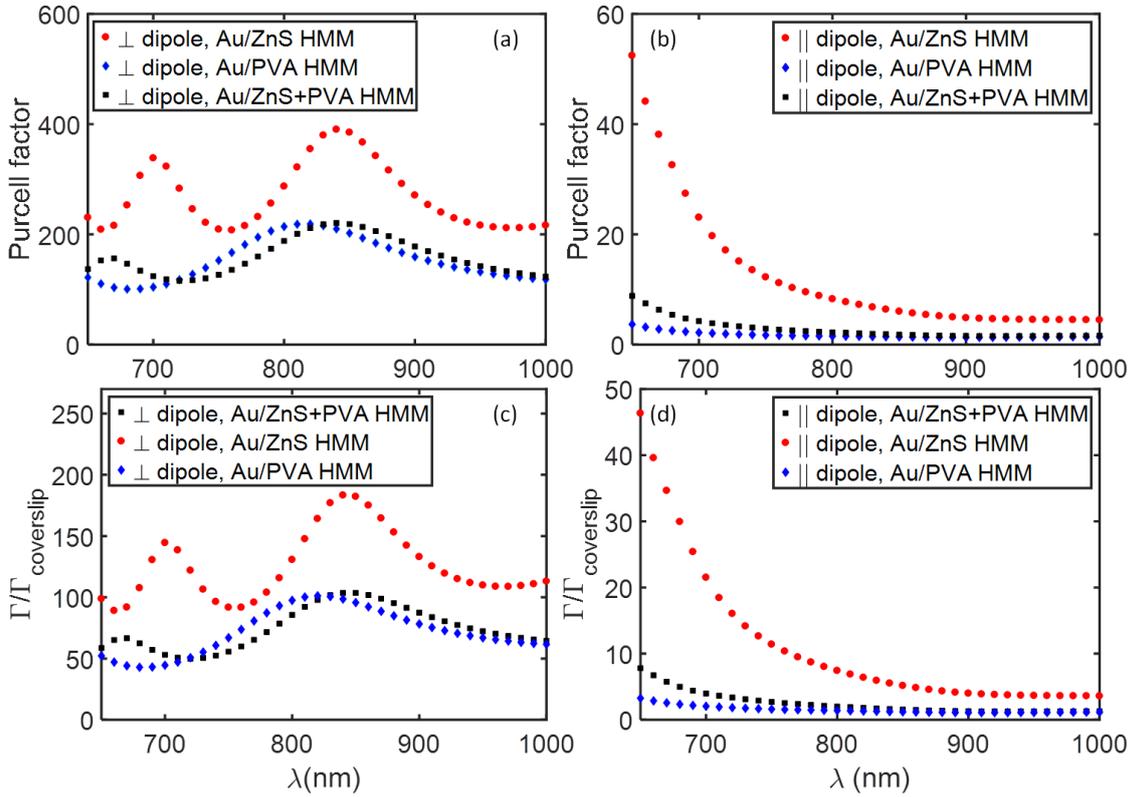

Figure 4. Purcell factor ($\Gamma_{HMM}/\Gamma_{vacuum}$) and relative emission rates ($\Gamma_{HMM}/\Gamma_{coverslip}$) for dipole emission in the above three HMM structures corresponding to (a), (c) perpendicular dipole orientation relative to the Au layers and (b), (d) parallel dipole orientation relative to the Au layer.

Figure 4 displays how the Purcell factor ($F_p = \Gamma_{HMM}/\Gamma_{vacuum}$) and relative emission rates ($\Gamma_{HMM}/\Gamma_{coverslip}$) for dipole emission in the HMM structure, gets significantly enhanced by the use of index-matched ZnS layers. On replacing the PVA layer ($n = 1.47$ at $\lambda = 900$ nm) with the high index ZnS layer ($n$ = 2.30 at λ = 900 nm) the enhancement in the Purcell factor/emission rate significantly increases. More prominent resonance peaks emerge in the emission spectrum.

These successive peaks results from the lossy Fabry-Perot structure modes [43]. The high index HMM structure itself acting a lossy Fabry-Perot cavity due to large reflections at the interfaces. Since the effective index of the medium decreases with increase in wavelength, the successive resonance peaks appears broader due to reduced reflectivity at the HMM interface. The broadband response is restricted to the case of dipole emission perpendicular to the Au interface. Perpendicular dipole emission experiences reflections at the top and bottom HMM layers. This results in the excitation of lossy Fabry-Perot modes in the HMM structure. The mixing of the plasmonic modes with these structure modes results in the overall broadband response from the structure. The resonance of these structure modes can be tuned across the infrared spectrum by controlling the metal and dielectric layer thicknesses.

Here we have chosen a thickness for the metal/dielectric layers for which the HMM structure could be easily fabricated. The main highlight of this study is to present a simple 5-layered HMM structure based on a high index dielectric layer having its index closely matched to the SiC and diamond based single photon emitters. This relatively simple strategy while preserving the hyperbolicity condition ($\epsilon_\perp \cdot \epsilon_\parallel < 0$) in our studied frequency range of 650 nm to 1000 nm (Fig. 1), results in a significantly large broadband enhancement over the full emission range of the various point emitters embedded inside the SiC and diamond nanoparticles. The chosen thicknesses of the metal/dielectric layers provide successive Fabry-Perot modes in the studied emission band of 650 nm – 1000 nm. With reduced thicknesses for metal/dielectric layers, more metal/dielectric layers would be required to achieve similar resonance modes in the studied emission range. This would have increased the complexity of the structure.

For the parallel dipole emission, the reflections at the top and bottom HMM layers are minimal due to the continuity of the field across the interface boundaries. The Purcell/emission enhancement is therefore restricted to only the plasmonic resonance in the visible spectrum without any structure modes arising in the infrared region (Fig. 4 (b)).

Here only the perpendicular dipole orientation gave an enhanced broadband response with successive resonance peaks in the infrared spectrum where the emission from SiC as well as diamond centres lies. However, as the spontaneous emission rate Γ of a dipole in the weak coupling regime from the Fermi's Golden rule is proportional to $\mu^2 \sum_{\mathbf{k},\sigma} |\hat{\mathbf{d}} \cdot \mathbf{E}_{\mathbf{k},\sigma}(\mathbf{r})|^2 \delta(\omega - \omega_{\mathbf{k},\sigma})$, with $\omega$ being the transition frequency, $\boldsymbol{\mu} = \mu \hat{\mathbf{d}}$ the dipole moment and $\mathbf{E}_{\mathbf{k},\sigma}(\mathbf{r})$ being the local electric field profile of the modes with wave-vectors $\mathbf{k}$ and polarizations $\sigma$ [40]. The Purcell enhancement is therefore expected to go as a function of $\cos^2\theta$, being about half for dipoles oriented at 45º to the interface. The Purcell enhancement factor being still more than 100 over the broad spectrum range of 650 nm to 1000 nm (refer to Figure 8a.).

To test the sensitivity in the Purcell/emission enhancement with the variations in the dipole positions within the SiC nanoparticles we varied the dipole position along the *z* (vertical) and the *x* (horizontal) directions. The *x* and *y* directions which are both parallel to the metal-dielectric interface are symmetric to each other. The HMM structure is considered to be continuous along the *x*, *y* (horizontal) directions and we have modelled these by using scattering boundary conditions along these directions. Figure 5 shows the variation in the Purcell factor with dipole displacement along the three orthogonal coordinate axes for the case of dipole emission at 900 nm, corresponding to vacancy-Si, $V_{Si}$ in SiC. Along the *z*-direction, as the dipole separation from the centre increases the dipole gets closer to the metal layers leading to significantly enhanced HMM structure modes coupled to the plasmonic resonance. Along the *x* and *y* directions, the dipole experiences the same electromagnetic environment due to the top and bottom HMM layer. Slight changes in the Purcell factor arise due to small variations in the dipole emission rates corresponding to the varying dipole separation from the SiC sphere surface for the vacuum case. Our result therefore shows that the enhancement is not specific to the dipole central location and the same order of enhancement exists for all dipole positions within the middle ZnS layer.

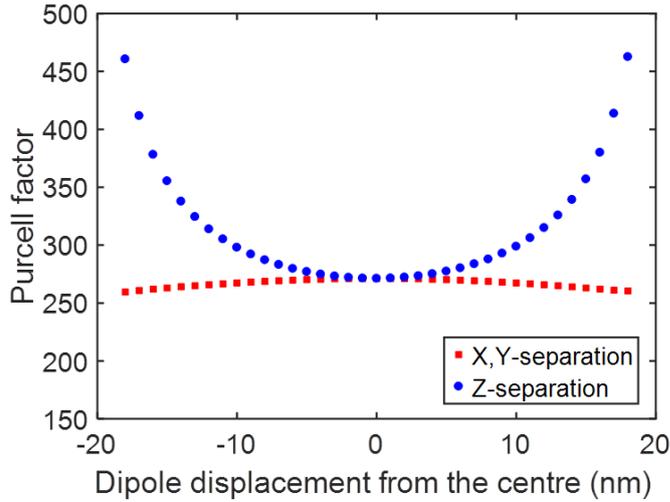

Figure 5. Variation of Purcell factor SiC based dipole emitter encapsulated in the multilayered Au/ZnS HMM structure with its displacement along the coordinate axes directions. The emission wavelength considered here is 900 nm corresponding to vacancy-Si, $V_{Si}$ in SiC.

**Collected Photon Rate Calculations:**

We have achieved broadband enhancement from visible to infra-red frequencies using a multilayered higher index dielectric HMM structure. We will now focus on how to capture the maximum light out of these structures. Figure 6 displays the far-field coupling of the dipole emission at 900 nm for a vertically oriented dipole radiating on a silica glass substrate ($n$ = 1.45) and in our HMM structure. Dipole emission on a glass substrate has been well studied with the glass itself being known to act as a dielectric antenna directing most of the emission to the higher index substrate medium [44,45]. For perpendicular dipole emission in the HMM, from Figure 6 it can be seen that most of the dipole emission is being lost in the HMM layer and the bottom glass substrate. We therefore need to design a robust antenna which would be able to direct this broadband Purcell enhancement into the numerical aperture (NA) of the collection objective on top of the HMM structure. This would result in a large photon collection rate (CPR) throughout the broadband spectral range of the HMM. For these photon collection rate calculations, we will consider the objective lens to have a numerical aperture of (NA) of 0.95. This corresponds to a collection solid angle of 71.8º.

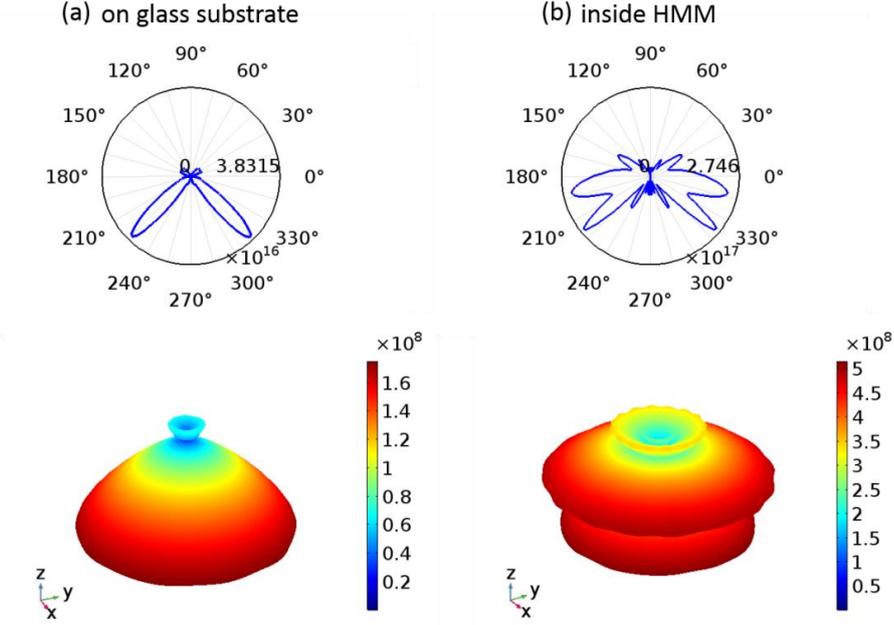

Figure 6. The far-field distribution of the radiated power for dipole emission in SiC nanoparticle when placed on a glass substrate and in the HMM structure. The emission wavelength considered is 900 nm.

To collect light from the top of HMM structures, complex periodic-gratings are commonly used [24,31]. Here we use a simple antenna structure consisting of a single metal cylinder on top of the HMM structure. Traditionally for radio frequencies, antenna parameters are prescribed only in terms of external wavelength. However, at optical frequencies electrons in the metal offer substantial inertia and do not respond instantaneously to the driving fields. Metal electrons therefore have to be treated as a strong coupled plasma at optical frequencies. This leads to a reduced effective wavelength within the antenna [46].

Figure 7a shows the schematic of the Au cylindrical antenna on top of the HMM structure. To optimize the antenna parameters for maximum collection efficiency we measured the collection efficiency (CE) from the structure by varying both the antenna height and antenna diameter at dipole emission wavelength equal to 900 nm. The calculated optimum antenna height and diameter for 900 nm emission wavelength are 110 nm and 820 nm respectively.

Figure 7b displays the far-field distribution of the radiated power for the dipole emission in SiC nanoparticle at 900 nm with Au cylindrical antenna on the HMM surface. The antenna is clearly able to direct a significant portion of the radiated power within the collection angle of the objective. Figure 8 shows the performance of this antenna for perpendicular and 45° to Au/ZnS interface dipole orientations in terms of the Purcell enhancement, the quantum efficiency (QE), the collection efficiency (CE) and the photon collected rate (CPR) of the HMM antenna structure. The quantum efficiency being defined as, $QE =$ far-field radiated power/total emitted power by the dipole.

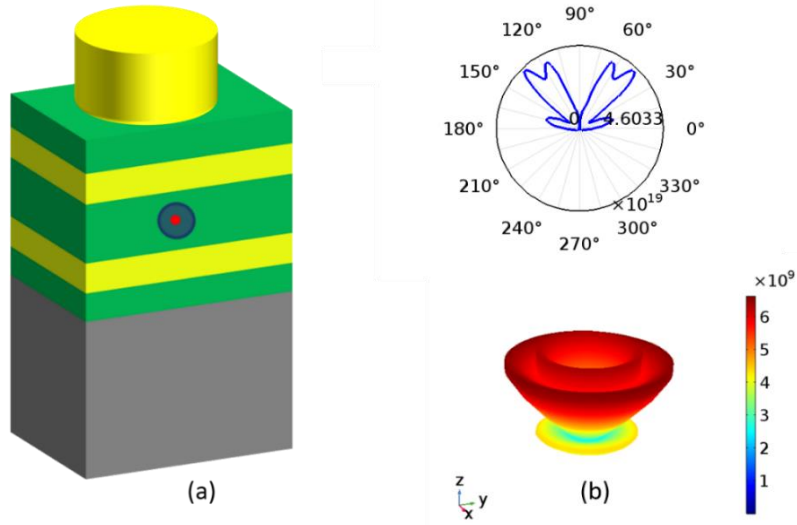

Figure 7. (a) Schematic of the Au cylindrical antenna on top of the HMM structure. (b) The far-field distribution of the radiated power for dipole emission in SiC nanoparticle when placed in the HMM structure with the cylindrical antenna on top. The emission wavelength considered is 900 nm.

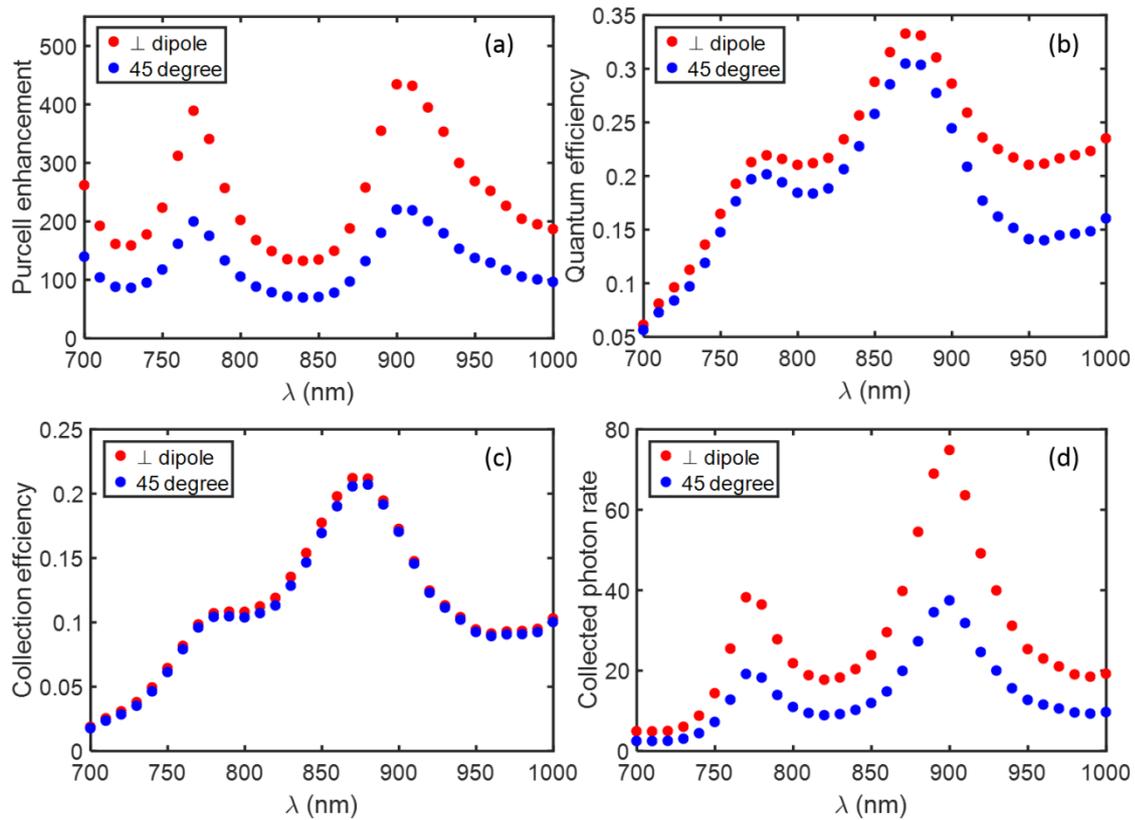

Figure 8. The total performance of the Au cylindrical antenna based HMM structure for perpendicular and 45º dipole orientation. (a) the modified Purcell enhancement ($F_p$), (b) the quantum efficiency (QE), (c) the collection efficiency and (d) the collected photon count rate (CPR).

The Au cylindrical antenna based HMM structure seems to shift the Purcell enhancement curve giving a resonance peak at 900 nm since we optimized the antenna parameters, its height and radius for dipole emission at 900 nm. However, the order of enhancement is well maintained throughout the broadband spectrum range of the various color centres in SiC and diamond. Though this simple design we are able to achieve an average collected photon rate (CPR) of about 30 throughout the broad spectral range of 700 nm to 1000 nm for perpendicular dipole orientation. Also since our Au antenna diameter is quite large, about 820 nm, the antenna performance will not be critical to the dipole location below the antenna and gives the same order of enhancement for few nanometers variation in the dipole positions within the HMM structure.

Since $CPR \propto F_p \cdot CE$ with $F_p \propto \cos^2 \theta$ and collection efficiency being mostly a function of the antenna only, the CPR is also expected to scale as $CPR \propto \cos^2 \theta$. From Figure 8, it can be seen that both Purcell enhancement and collected photon rates are nearly being halved for 45º dipole orientation relative to the Au/ZnS interface. The collection efficiency is similar for both the dipole orientations. The 45º dipole orientation also gives a high average CPR value of about 15 over the broad spectral range of 700 nm to 1000 nm.

The antenna design here is optimized for dipole emission around 900 nm corresponding to SiC nanoparticles based color centres. However, this scheme can be applied to optimize the antenna performance across any of the color centres in diamond and SiC, including the well-known NV or Si vacancy centres by tuning the antenna parameters (its height and radius).

## IV. CONCLUSIONS:

We have demonstrated a simple HMM structure based on a high refarctive index dielectric, ZnS, and Au layers which shows a large broadband Purcell enhancement for dipole emission in both visible and near infra-red regions. The HMM resonator is composed of gold and ZnS layers to achieve a Purcell enhacement of 400 at 850 nm and 300 at 680nm with a similar order of enhancement throughout the broad spectral range of 650 nm to 1000 nm.

By employing simple gold (Au) cylindrical antenna on top of the HMM structure we are able to achieve a large average collected photon rate (CPR) of about 30 throughout the broad spectral range of 700 nm to 1000 nm. The peak CPR value of about 80 at 900 nm corresponding to the emission of silicon-carbide quantum emitters. This is therefore a state-of-the art improvement considering that the previous computational design using a hybrid bowtie metal-dielectric antenna has reported a maximum average CPR of 25 across the NV emission spectrum of 600 nm to 800 nm with the highest value being about 40 at 650 nm [27]. In this calculation, the NV dipole is considered to be located at the centre of the bowtie with its orientation along the bowtie axis.

As the refractive indices of ZnS ($n$ = 2.30 at 900 nm), SiC (2.59) and diamond (2.39) are matched, our HMM scheme is universal for all SiC and diamond based emitters. It will lead to similar enhancements for the case of SiC nanocrystals fabricated as in [33,47] and embedded in the ZnS middle layer; or for nanodiamond embedded in the ZnS middle layer. Additionally, this model is particularly interesting as it can be applied to enhance emission of single photon sources not only in nanoparticles emitters but also in thin layer. This design can be extended to the case of emitters embedded in diamond [48] or SiC thin-films by replacing the middle ZnS layer.

By replacing the dielectric with SiC itself and the metamaterial with TiN (which has similar plasmonic properties of gold) a fully CMOS compatible HMM resonator can be realised. Due to TiN being refractory material, its fabrication is compatible with thin film SiC pulsed laser deposition [38] or SiC chemical vapor deposition, where surface color centres with dipole aligned to the main crystallographic axis can be created [49] in addition to the above mentioned color centres.

## V. METHODS:

All electromagnetic calculations of the dipole emission rate and collected power are performed using the commercial finite-element method (FEM) based COMSOL Multiphysics RF module package 5.2. The single vacancy color center is modelled as an oscillating point dipole [50] located within the SiC nanoparticle of size 40 nm.

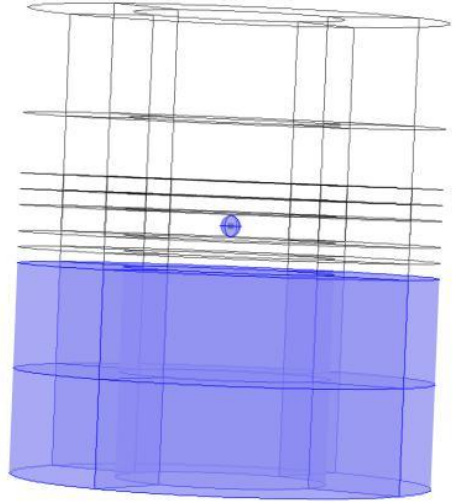

Figure 9. Computational geometry of our HMM structure. The highlighted bottom domains corresponds to the coverglass surface on which our HMM structure is place. The inner sphere corresponds to the SiC nanoparticle enclosing the point dipole.

Spontaneous emission is a purely quantum process. However since the influence of the environment is expressed through the classical LDOS, the emission rate relative to a reference system can be found by classical electromagnetic calculations also [50]. Here in this study we are accounting for only the radiative emission process considering our emitters to have high quantum efficiency. Treating the quantum emitter as a classical point dipole, the electromagnetic fields are excited by a point current source driven at frequency $v = c/\lambda$. The total power radiated by the dipole is calculated over the surface of the SiC sphere as highlighted in Figure 9. The radiative spontaneous emission rate enhancement is then calculated as $R = P/P_r$, where $P_r$ is the power corresponding to the reference system. For Purcell factor calculations the reference system is a SiC nanoparticle in vacuum. Within the whole computational domain the minimum mesh size used is 0.5 nm and the maximum mesh size used is 45 nm. Within the SiC sphere where the radiated power is calculated the minimum mesh size is kept at 0.5 nm with the maximum mesh size being 2 nm.

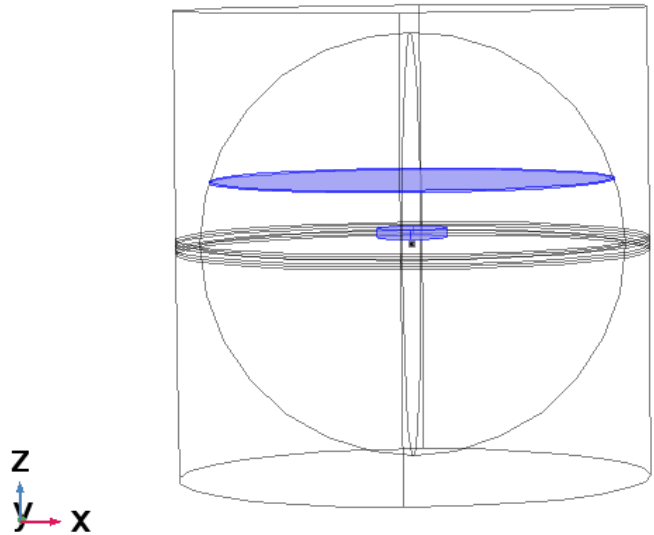

Figure 10. Computational geometry of our HMM antenna structure for far-field and collected photon rate calculations.

Figure 10 shows the computational geometry of our HMM structure with the radiated power being collected on the top highlighted surface. The diameter of the top surface is chosen such that it acts as a collection lens with a numerical aperture of 0.95. The far-field power distribution is calculated on the surface of the enclosed sphere whose radius is about 3 times the dipole emission wavelength. The Au cylindrical antenna is represented by the highlighted cylindrical surface on top of the HMM structure.

**Acknowledgement**

FAI would like to thank Venu Gopal Achanta, TIFR, India, for useful discussions during the course of this study. FAI acknowledge University Grant Commission, India for funding through the Faculty Start-up grant.